\documentclass[prd,twocolumn,twoside,preprintnumbers,superscriptaddress,nofootinbib]{revtex4}

\usepackage{amsmath,slashed}
\usepackage{graphicx,graphics,color}
\usepackage{dcolumn}
\usepackage[hyperfootnotes=false]{hyperref}
\usepackage{xspace}
\usepackage{enumerate}
\usepackage{varwidth}

\usepackage{physics}
\usepackage{amssymb}
\usepackage{mathtools}
\usepackage{dsfont}
\usepackage{multirow}

\newcommand{\GeV}{\,\text{GeV}}
\newcommand{\MeV}{\,\text{MeV}}

\newcommand{\mpi}{M_\pi}
\newcommand{\Fpi}{F_\pi}
\newcommand{\mpic}{M_{\pi^+}}
\newcommand{\mpin}{M_{\pi^0}}

\newcommand{\eps}{\epsilon}
\renewcommand{\Im}{\text{Im}\,}
\renewcommand{\Re}{\text{Re}\,}
\newcommand{\Order}{\mathcal{O}}
\newcommand{\beq}{\begin{equation}}
\newcommand{\eeq}{\end{equation}}
\newcommand{\Lagr}{\mathcal L}

\allowdisplaybreaks[1]

\begin{document}

\preprint{PSI-PR-22-33, ZU-TH 55/22}

\title{Width effects of broad new resonances in loop observables and application to $\boldsymbol{(g-2)_\mu}$}

\author{Andreas Crivellin}
\affiliation{Paul Scherrer Institut, 5232 Villigen PSI, Switzerland}
\affiliation{Physik-Institut, Universit\"at Z\"urich, Winterthurerstrasse 190, 8057 Z\"urich, Switzerland}
\author{Martin Hoferichter}
\affiliation{Albert Einstein Center for Fundamental Physics, Institute for Theoretical Physics, University of Bern, Sidlerstrasse 5, 3012 Bern, Switzerland}

\begin{abstract} 
In the phenomenology of strong interactions most physical states acquire a substantial width, and thus can only be defined in a model-independent way by pole positions and residues of the $S$-matrix. This information is incorporated in the K\"all\'en--Lehmann representation, whose spectral function characterizes  the shape of the resonance and can be constrained by the dominant decay channels. Here, we argue that similar effects become important whenever beyond-the-Standard-Model particles possess a sizable decay width---as possible for instance in cases with a large branching fraction to a dark sector or strongly coupled scenarios---and show how their widths can be incorporated in the calculation of loop observables. 
 As an application, we consider the anomalous magnetic moment of the muon, including both the direct effect of new physics and the possible indirect impact of a broad light $Z'$ on $e^+e^-\to\text{hadrons}$ cross sections. Throughout, we provide results for a general spectral function and its reconstruction from the one-loop imaginary part, where the latter captures the leading two-loop effects.   
\end{abstract}

\maketitle

\section{Introduction}

While the existence of physics beyond the Standard Model (BSM) is established by experimental observations, in particular of neutrino masses and dark matter at cosmological scales, the properties of the required new particles and interactions remain unclear. Therefore, these observations are insufficient to construct  the fundamental theory superseding the SM. Anomalies in precision experiments, see, e.g., Refs.~\cite{Crivellin:2021sff,Crivellin:2022qcj,Fischer:2021sqw} for recent reviews, suggest certain patterns for the novel interactions, most notably the violation of lepton flavor universality, but only the ratio of couplings over masses can be accessed and the widths of the new states remain elusive. Nonetheless, for a given fixed effect in low-energy observables, the couplings must increase with the mass, leading unavoidably to larger widths of the new states. In fact, bounds on the couplings of new particles from perturbativity and unitarity have been derived in this context~\cite{DiLuzio:2017chi,Capdevilla:2021rwo,Allwicher:2021jkr}. Importantly, even below these bounds the width of the new particles is sizable and such large widths were used in several models to avoid or weaken collider bounds~\cite{Buttazzo:2017ixm,Iguro:2020ndk,Calibbi:2017qbu}, in particular by decays to invisible final states~\cite{Sala:2017ihs,Mohapatra:2021izl,Datta:2017ezo,Altmannshofer:2017bsz,Sala:2018ukk,Bishara:2017pje,Borah:2020swo,Darme:2021qzw,Greljo:2021npi,Crivellin:2022obd,Alok:2022pjb,Datta:2022zng}. 
Examples include 
$e^+e^-\to \mu^+\mu^-+\text{invisible}$, $e^+e^-\to\gamma+\text{invisible}$ searches~\cite{Adachi:2019otg} as well as mono-photon and mono-jet searches at the LHC~\cite{ATLAS:2021kxv,CMS:2021far,CMS:2017qyo}. Therefore, the question arises how to properly include broad width effects---in particular in strongly coupled regimes, as arise naturally in composite~\cite{Farhi:1980xs,Dine:1981za} or extra-dimensional~\cite{Randall:1999ee,Antoniadis:1990ew,ArkaniHamed:1998kx} models---into the calculation of low-energy observables. In fact, this problem has, to the best of our knowledge, so far not been addressed in the context of BSM physics.

\begin{figure}[t]
 \includegraphics[width=0.8\linewidth]{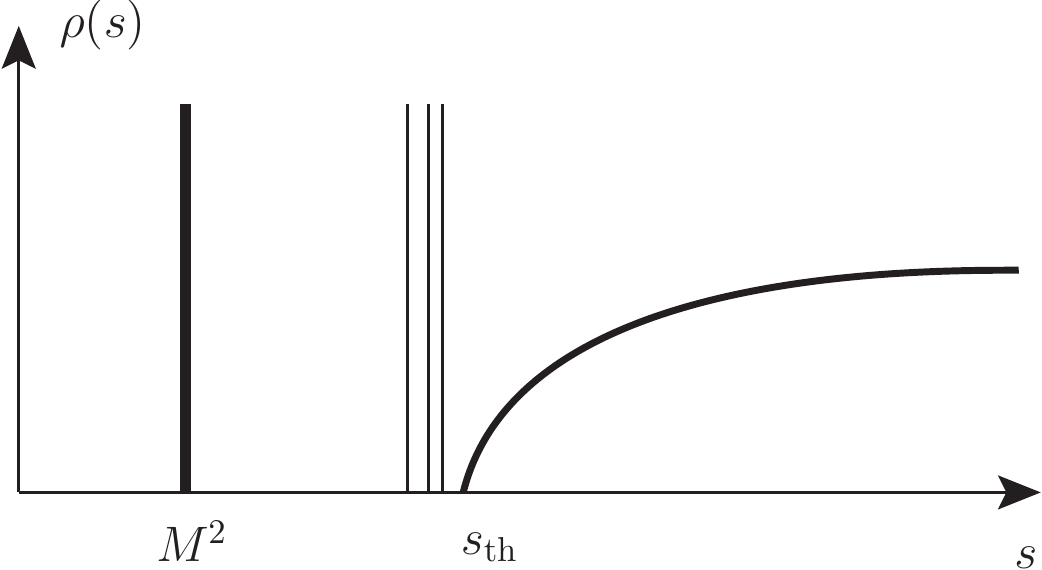}
 \caption{Generic form of the KL spectral function $\rho(s)$. Single-particle poles show up as $\delta$ functions at the respective mass, $s=M^2$, while multi-particle states lead to a continuum that starts at some threshold $s_\text{th}$, below which bound states are possible (indicated by the narrow lines).}
 \label{fig:KL}
\end{figure}

Here we argue that this outstanding problem can be solved via the application of the K\"all\'en--Lehmann (KL) spectral representation~\cite{Kallen:1952zz,Lehmann:1954xi}, which describes the general form of a time-ordered two-point function of an interacting quantum field theory. It can be derived on general grounds by inserting a complete set of states and using Lorentz invariance, with a result that makes the analytic structure of the two-point function manifest. The KL representation for a scalar particle is given by
\beq
\label{KL}
\Delta_\phi(p^2)=\int_0^\infty ds\, \frac{\rho_\phi(s)}{p^2-s+i\eps},
\eeq
where the propagator of the free theory with squared mass $s$ is convolved with a general spectral density $\rho_\phi(s)$ (subject to the normalization condition $\int ds\,\rho_\phi(s)=1$ and positivity constraint $\rho_\phi(s)\geq 0$). A one-particle state corresponds to an isolated pole in $\rho_\phi(s)$, multi-particle states to a branch cut starting at the respective threshold $s_\text{th}$, and bound states  to additional poles below $s_\text{th}$, see Fig.~\ref{fig:KL}.   
While the existence of such spectral representations is a general non-perturbative result, a practical application concerns the incorporation of width effects in the phenomenology of strong interactions, e.g., by improving Breit--Wigner (BW) parameterizations~\cite{Breit:1936zzb} through variants with good analytic properties~\cite{Lomon:2012pn,Moussallam:2013una,Zanke:2021wiq}, which can then be used in loop calculations without generating spurious imaginary parts. As an added benefit, the denominator of the KL representation can be efficiently integrated in higher-order computations as it resembles the standard Feynman propagator with $s=M^2$, so that merely the final result has to be convolved with the spectral function.

In general, once a resonance is located deep in the complex plane---the $f_0(500)$~\cite{Caprini:2005zr,Hoferichter:2011wk,Garcia-Martin:2011nna,Moussallam:2011zg,Pelaez:2015qba} and the $K_0^*(700)$~\cite{Descotes-Genon:2006sdr,Pelaez:2020gnd} being typical examples---only a description in terms of the pole position and its residues is viable. However, for moderate widths---such as the $\rho(770)$, with $\Gamma_\rho/M_\rho\simeq 20\%$---a KL representation with spectral function derived from the main decay channels (potentially supplemental by centrifugal-barrier factors~\cite{COMPASS:2015gxz,VonHippel:1972fg} to dampen the high-energy behavior) is often phenomenologically successful, with resonance properties close to the actual pole parameters~\cite{Colangelo:2001df,Garcia-Martin:2011nna,Hoferichter:2017ftn}. In particular, representations of the form~\eqref{KL} can be used to implement resonance effects in loop observables without introducing unphysical imaginary parts below the respective thresholds~\cite{Zanke:2021wiq,Colangelo:2022lzg}. 

The main point of this article is that a similar strategy can also be applied to physics beyond the SM, whenever the new particles acquire a sizable width. In general, the results are formulated in terms of a spectral function that needs to be determined from experiment---a prime example in the SM being hadronic vacuum polarization (HVP). In the absence of data, the spectral function can still be constrained from the perturbative imaginary part once a given decay channel is assumed, especially, the threshold behavior and the functional form in the vicinity of the resonance, and systematic improvements are possible by going to subleading orders in perturbation theory~\cite{Sirlin:1991fd,Passera:1996nk,Passera:1998uj}.     
In the following, we will first present the general formalism, including the explicit form of the one-loop spectral functions for different quantum numbers of all particles involved, before turning to the application to the exemplary case of the anomalous magnetic moment of the muon.

\section{Spectral functions and K\"all\'en--Lehmann representation}
\label{sec:KL}

While the form of the KL representation in Eq.~\eqref{KL} holds in general for a scalar particle, for practical applications the spectral function is required. In principle, in the case of a broad resonance its form can be extracted experimentally, e.g., from scattering processes involving the decay products that give rise to the continuum in Fig.~\ref{fig:KL}, but already for hadronic reactions a complete measurement of spectral functions is complicated. However, useful approximations to $\rho(s)$ can be obtained for instance by analytically improved versions of BW parameterizations~\cite{Lomon:2012pn,Moussallam:2013una,Zanke:2021wiq}, in which case the energy dependence can be constrained via the imaginary part of the self energy, matching the imaginary part of Eq.~\eqref{KL} to the imaginary part of the resummed Dyson series. The result for the decay of a scalar $\phi$ with mass $M$, derived in App.~\ref{app:KL}, can be written in the form
\begin{align}
\label{rho_scalar_final}
 \rho_\phi(s)&=\frac{Z}{\pi}\frac{\sqrt{s}\,\Gamma(s)}{\big(s-M^2\big)^2+s \big[\Gamma(s)\big]^2},\notag\\
 \Gamma(s)&=\sum_{K=\phi_1\phi_2,F_1F_2}\Gamma_{\phi\to K}\frac{\gamma_{\phi\to K}(s)}{\gamma_{\phi\to K}(M^2)}\theta(s-s_K),
\end{align}
with total width $\Gamma=\Gamma_{\phi\to\phi_1\phi_2}+\Gamma_{\phi\to F_1 F_2}$. The energy dependence is described by
\begin{align}
\gamma_{\phi\to\phi_1\phi_2}(s)&=\frac{\lambda^{1/2}\big(s,M_1^2,M_2^2\big)}{s^{3/2}},\notag\\
\gamma_{\phi\to F_1F_2}(s)&=\frac{\lambda^{1/2}\big(s,m_1^2,m_2^2\big)}{s^{3/2}}\notag\\
&\times\Big(s-m_1^2-m_2^2-2\xi_\phi m_1 m_2\Big),
\end{align}
with the K\"all\'en function $\lambda(a,b,c)=a^2+b^2+c^2-2(a b+a c+bc)$, masses $M_i$ ($m_i$) for the scalar (fermionic) decay products $\phi_i$ ($F_i$), and a parameter $\xi_\phi=\big[(C_S^\phi)^2-(C_P^\phi)^2\big]/\big[(C_S^\phi)^2+(C_P^\phi)^2\big]\in[-1,1]$ that describes the chirality of the couplings, see App.~\ref{app:self}. The thresholds are $s_{\phi_1\phi_2}=(M_1+M_2)^2$, $s_{F_1 F_2}=(m_1+m_2)^2$. Finally, the parameter $Z$ is to be determined from the normalization condition $\int ds\, \rho_\phi(s)=1$, and Eq.~\eqref{rho_scalar_final} generalizes accordingly when further decay channels contribute. In the form of Eq.~\eqref{rho_scalar_final}, mass and width of the resonance (together with branching fractions and masses of the decay products) need to be provided as input. If the width is sizable but not too large, it can also be calculated perturbatively, see Eq.~\eqref{Gamma_pert}, in which case inserting the KL representation into a one-loop diagram captures the leading two-loop effect, and similarly at subleading orders~\cite{Sirlin:1991fd,Passera:1996nk,Passera:1998uj}. 

For non-chiral fermions the KL representation generalizes to~\cite{Itzykson:1980rh}
\beq
\label{KL12}
\Delta_F(p)=\int_0^\infty ds\, \frac{\slashed{p}\rho_1(s)+\rho_2(s)}{p^2-s+i\eps},
\eeq
involving two spectral functions with positivity conditions for $\rho_1(s)$ and $\sqrt{s}\rho_1(s)-\rho_2(s)$ as well as the normalization $\int ds \, \rho_1(s)=1$. The determination of the spectral function in terms of the imaginary part of the self energy is accurate up to terms of second order $\Order\big((s-m^2)^2\big)$ in the expansion around the resonance mass $m$. At the same level of precision we find that  
 $\rho_1(s)=\rho_2(s)/\sqrt{s}\equiv \rho_F(s)$, while the generalization to chiral fermions, given in App.~\ref{app:KL}, can potentially introduce a first-order correction. The result for a broad fermion $F$ with decays to fermion ($F'$) and scalar ($\phi'$) or vector ($X'$) becomes
 \begin{align}
 \label{rho_12_final}
  \rho_F(s)&=\frac{Z}{\pi}\frac{m \Gamma(s)}{(s-m^2)^2+m \sqrt{s}\big[\Gamma(s)\big]^2},\notag\\
 \Gamma(s)&=\sum_{K=F'\phi',F'X'}\Gamma_{F\to K}\frac{\gamma_{F\to K}(s)}{\gamma_{F\to K}(m^2)}\theta(s-s_K),
\end{align}
with total width $\Gamma=\Gamma_{F\to F'\phi'}+\Gamma_{F\to F' X'}$, thresholds $s_{F'\phi'}=(m_{F'}+M_{\phi'})^2$, $s_{F' X'}=(m_{F'}+M_{X'})^2$, and energy dependence 
\begin{align}
\gamma_{F\to F'\phi'}(s)&=\frac{\lambda^{1/2}\big(s,m_{F'}^2,M_{\phi'}^2\big)}{s^{3/2}}\\
&\times\Big(s+m_{F'}^2-M_{\phi'}^2+2\xi_{\phi'}m_{F'}\sqrt{s}\Big),\notag\\
\gamma_{F\to F'X'}(s)&=\frac{\lambda^{1/2}\big(s,m_{F'}^2,M_{X'}^2\big)}{s^{3/2}}\bigg[\lambda\big(s,m_{F'}^2,M_{X'}^2\big)\notag\\
&+3M_{X'}^2\Big(s+m_{F'}^2-M_{X'}^2-2\xi_{X'}m_{F'}\sqrt{s}\Big)\bigg],\notag
\end{align}
where $\xi_{\phi'}=(C_S'^2-C_P'^2)/(C_S'^2+C_P'^2)$, $\xi_{X'}=(C_V'^2-C_A'^2)/(C_V'^2+C_A'^2)$ again determine the chirality of the couplings (see App.~\ref{app:self}) and $Z$ follows from the normalization $\int ds \, \rho_F(s)=1$.

Finally, the generalization of the KL representation to the spin-$1$ case becomes complicated by the presence of unphysical degrees of freedom in the covariant formulation, and thus the need to specify the choice of gauge~\cite{Itzykson:1980rh}. 
In Feynman gauge, the Goldstone part can be evaluated using Eq.~\eqref{rho_scalar_final}, while the spectral function for the transverse component 
\beq
\Delta_X^{\mu\nu}(p)=g^{\mu\nu}\int_0^\infty ds\frac{\rho_X(s)}{p^2-s+i\eps}
\eeq
takes the form
\begin{align}
\label{rho_spin_1}
 \rho_X(s)&=\frac{Z}{\pi}\frac{\sqrt{s}\Gamma(s)}{(s-M_X^2)^2+s\big[\Gamma(s)\big]^2},\notag\\
 \Gamma(s)&=\sum_{K=\phi_1\phi_2,F_1F_2}\Gamma_{X\to K}\frac{\gamma_{X\to K}(s)}{\gamma_{X\to K}(M_X^2)}\theta(s-s_K),
\end{align}
with
\begin{align}
\label{gamma_spin_1}
\gamma_{X\to\phi_1\phi_2}(s)&=\frac{\lambda^{3/2}\big(s,M_1^2,M_2^2\big)}{s^{5/2}},\notag\\
\gamma_{X\to F_1F_2}(s)&=\frac{\lambda^{1/2}\big(s,m_1^2,m_2^2\big)}{s^{5/2}}\bigg[-\lambda\big(s,m_1^2,m_2^2\big)\notag\\
&+3s\Big(s-m_1^2-m_2^2+2m_1m_2\xi_X\Big)\bigg],
\end{align}
and $\xi_X=\big[(C_V^X)^2-(C_A^X)^2\big]/\big[(C_V^X)^2+(C_A^X)^2\big]$. 

In the application of the preceding expressions in loop integrals, one additional subtlety concerns the high-energy behavior. To ensure the required decoupling limit for large momenta, Eq.~\eqref{KL} needs to behave as $\Delta_\phi(p^2)\sim 1/p^2$, which reproduces the normalization condition $\int ds\,\rho_\phi(s)=1$ as a superconvergence  relation~\cite{Martin:1970hmp}. However, the exchange of limits here is delicate, and in general subtractions may be necessary~\cite{Weinberg:1995mt}. Given that our derivation of the spectral functions $\rho_\phi$, $\rho_F$, $\rho_X$ is only accurate up to terms $\Order\big((s-M^2)^2\big)$ in the first place, a convenient way to ensure convergence is based on the observation that multiplication by the factor
\beq
\label{xis}
\big[\xi(s)\big]^n=\left(\frac{2M\sqrt{s}}{s+M^2}\right)^n=1+\Order\big((s-M^2)^2\big),
\eeq
changes the high-energy behavior, without affecting the resonance physics at the claimed accuracy. Similar relations have already been used in the derivation of Eq.~\eqref{rho_12_final}, see App.~\ref{app:KL},  so that in this case the superconvergence relation is already well defined, while for the bosonic cases $n=1$ is sufficient to ensure convergence. Physically, the necessity to introduce modifications of the high-energy behavior as in Eq.~\eqref{xis} originates from the momentum factors associated with higher spins. In hadronic physics, 
the same effect can be achieved by adding centrifugal-barrier factors~\cite{COMPASS:2015gxz,VonHippel:1972fg}, but, in either case, ultimately 
the behavior of the spectral function off the resonance needs to be extracted from experiment.

\begin{figure}[t]
 \includegraphics[width=0.99\linewidth]{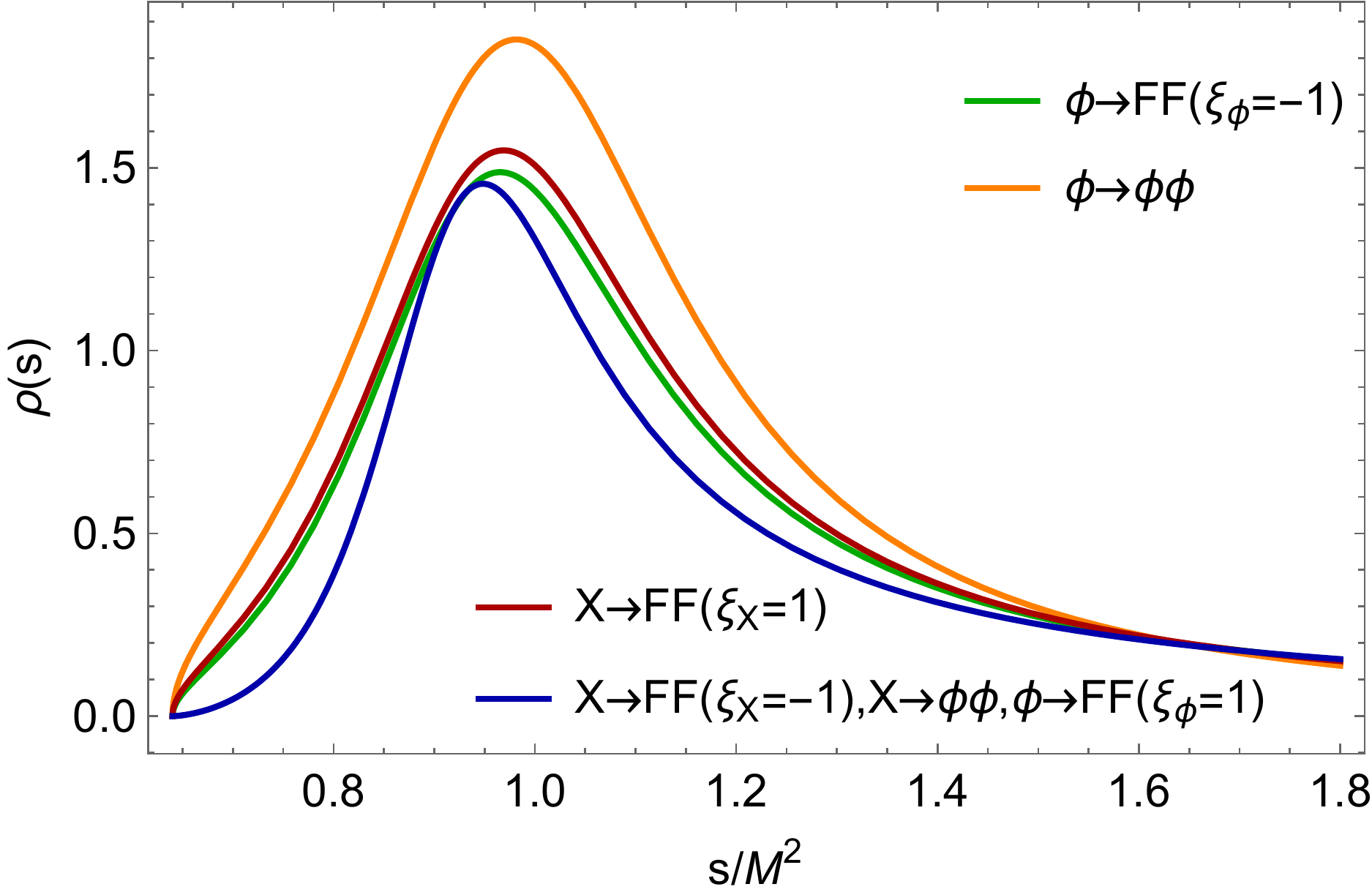}
 \caption{Spectral functions for the decay of scalars and spin-$1$ particles to a pair of fermions or scalars with mass $m/M=0.4$ and a width $\Gamma/M=0.2$ for $\xi_{X,\phi}=\pm1$.}
 \label{fig:spectralfuctions}
\end{figure}

The expressions presented here determine the amount of information that can be gleaned from resumming the $1$-loop self-energy diagrams, most importantly, the correct threshold behavior of the spectral function. We illustrate the spectral functions for scalars and spin-$1$ particles for the minimal integer $n$ required to achieve convergence in Fig.~\ref{fig:spectralfuctions}, but emphasize that the formalism is much more general, and becomes most powerful in cases in which the spectral function can be constrained from experiment. 

\section{Anomalous magnetic moment of the muon}
\label{sec:applications}

Broad new states will also leave their imprint 
in low-energy precision observables. As a specific and very relevant example we consider here the anomalous magnetic moment of the muon $a_\mu$ (see, e.g., Refs.~\cite{Jackiw:1972jz,Leveille:1977rc,Moore:1984eg,Kuno:1999jp,Cheung:2001ip,McKeen:2009ny,Jegerlehner:2009ry,Queiroz:2014zfa,Lindner:2016bgg,Crivellin:2017dsk,Crivellin:2018qmi,Crivellin:2021rbq,Athron:2021iuf} for the analysis of generic BSM scenarios). In App.~\ref{app:amu} we give the expressions for BSM effects in $a_\mu$~\cite{Leveille:1977rc}, generalized to the case of sizable widths using the KL representation. Depending on the size of the width such effects could significantly alter the parameter space required to explain the current $4.2\sigma$
discrepancy between experiment~\cite{Muong-2:2006rrc,Muong-2:2021ojo,Muong-2:2021ovs,Muong-2:2021xzz,Muong-2:2021vma} and the prediction in the SM~\cite{Aoyama:2020ynm,Aoyama:2012wk,Aoyama:2019ryr,Czarnecki:2002nt,Gnendiger:2013pva,Davier:2017zfy,Keshavarzi:2018mgv,Colangelo:2018mtw,Hoferichter:2019gzf,Davier:2019can,Keshavarzi:2019abf,Hoid:2020xjs,Kurz:2014wya,Melnikov:2003xd,Masjuan:2017tvw,Colangelo:2017qdm,Colangelo:2017fiz,Hoferichter:2018dmo,Hoferichter:2018kwz,Gerardin:2019vio,Bijnens:2019ghy,Colangelo:2019lpu,Colangelo:2019uex,Blum:2019ugy,Colangelo:2014qya} when HVPis derived from $e^+e^-\to\text{hadrons}$ cross-section data. 
In fact, in this data-driven evaluation a KL representation is used to implement, in a model-independent way, the effect of hadronic resonances. 

 The leading-order HVP contribution to $a_\mu$ can be represented by the master formula~\cite{Bouchiat:1961lbg,Brodsky:1967sr}
\begin{align}
\label{amu_HVP}
 a_\mu^\text{HVP}&=\bigg(\frac{\alpha m_\mu}{3\pi}\bigg)^2\int_{s_\text{th}}^\infty ds \frac{\hat K(s)}{s^2}R_\text{had}(s),\notag\\
 R_\text{had}(s)&=\frac{3s}{4\pi\alpha^2}\sigma(e^+e^-\to\text{hadrons}(+\gamma)),
\end{align}
where $\hat K(s)$ is an analytically known kernel function and the integration threshold $s_\text{th}=M_{\pi^0}^2$ is determined by the $\pi^0\gamma$ channel. Its derivation starts from a dispersion relation for the subtracted vacuum polarization function
\beq
\label{barPi_disp}
\bar \Pi(k^2)=\frac{k^2}{\pi}\int_{s_\text{th}}^\infty ds\frac{\Im \Pi(s)}{s(s-k^2)},
\eeq
which amounts to a KL representation for the two-point function of two electromagnetic currents with a spectral function determined by
the imaginary part
\beq
\label{ImPi_R}
\Im \Pi(s)=-\frac{\alpha}{3}R_\text{had}(s).
\eeq
The kernel function $\hat K(s)$ is obtained by performing the Feynman-parameter integral in $a_\mu$, but the derivation via Eq.~\eqref{barPi_disp} also admits a space-like master formula
\beq
\label{amu_space_like}
a_\mu^\text{HVP}=\frac{\alpha}{\pi}\int_0^1 dx(1-x)\bar \Pi(s_x),\qquad s_x=-\frac{x^2m_\mu^2}{1-x}. 
\eeq
In this form, Eq.~\eqref{amu_space_like} can be formally evaluated for any polarization function $\bar\Pi(s)$ regardless of its analytic properties without running into obvious inconsistencies, but the numerical result can be altered dramatically. As an example, the calculation of the $\pi^0\gamma$ contribution from Ref.~\cite{Blokland:2001pb} using an asymptotic expansion misses the correct value by a factor $10$, which can be traced back to the assumed form of  $\bar\Pi(s)$ that does not fulfill the dispersion relation~\eqref{barPi_disp}. This illustrates the importance of working with a representation of the two-point function with good analytic properties, see App.~\ref{app:pi0gamma} for a more detailed analysis. In particular, the $\pi^0\gamma$ example shows that the mistake incurred when violating analyticity properties is not always suppressed by the width of the states, since in this case the correct imaginary part actually scales with the inverse of the small width of the $\omega$, which ultimately produces the sizable enhancement of the $\pi^0\gamma$ contribution.

\begin{figure}[t]
 \includegraphics[width=0.99\linewidth]{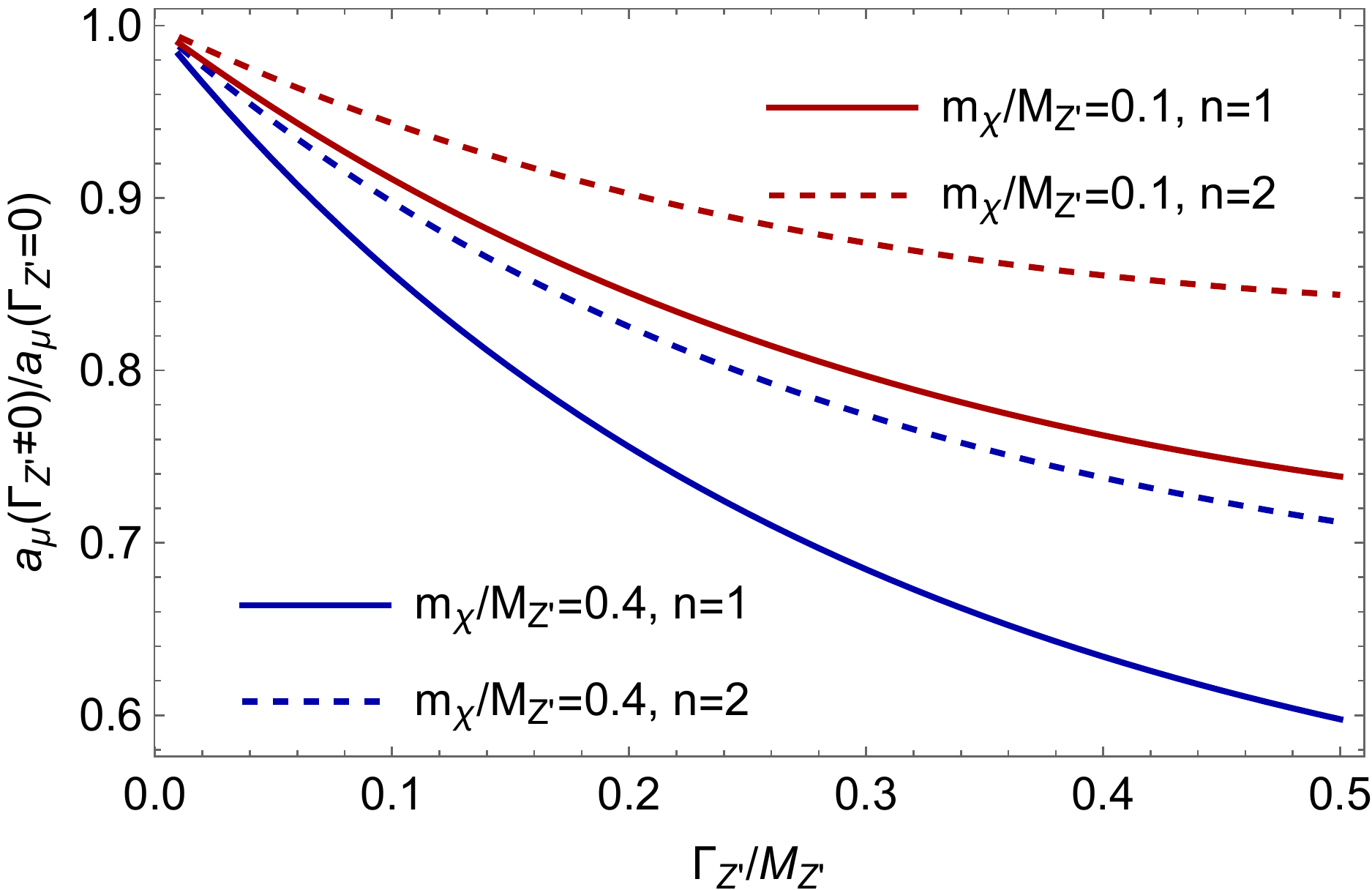}
 \caption{Relative effect of taking into account a non-zero width, compared to the case in which it is neglected, of a $Z^\prime$ boson in $a_\mu$ for $M_{Z^\prime}=1\GeV$ as a function of $\Gamma_{Z'}/M_{Z^\prime}$ for different powers ($n=1,2$) in  Eq.~\eqref{xis}.}
 \label{fig:amuZp}
\end{figure}

A simple example for a broad new state in $a_\mu$ is a neutral gauge boson ($Z'$). In particular, an $L_\mu-L_\tau$ symmetry~\cite{He:1990pn,Foot:1990mn,He:1991qd,Heeck:2011wj} constitutes an anomaly-free extension of the SM and is known to be capable of explaining the tension in $a_\mu$~\cite{Gninenko:2001hx,Baek:2001kca,Carone:2013uh,Altmannshofer:2014pba} while avoiding the bounds related to electrons (e.g., from $(g_2)_e$ and $e^+e^-\to e^+e^-$). Furthermore, it can serve as a portal to dark matter~\cite{Cirelli:2008pk,Baek:2008nz,Foldenauer:2018zrz,Okada:2019sbb,Holst:2021lzm,Drees:2021rsg,Heeck:2022znj}, which at the same time weakens collider bounds by introducing decays into invisible final states. In fact, a $Z^\prime$ with such a sizable invisible width was also studied in the literature as a possible solution to the $b\to s\ell^+\ell^-$ anomalies~\cite{Sala:2017ihs,Mohapatra:2021izl,Datta:2017ezo,Altmannshofer:2017bsz,Sala:2018ukk,Bishara:2017pje,Borah:2020swo,Darme:2021qzw,Greljo:2021npi,Crivellin:2022obd,Alok:2022pjb,Datta:2022zng}.
We therefore consider
\begin{align}
\Lagr_{Z^\prime}&=C_X  g^\prime\bar\chi\gamma^\mu\chi Z^\prime_\mu+\text{h.c.},
\end{align}
with a generic mass $M_{Z^\prime}$, as an example to illustrate the impact of a large width.\footnote{While we are assuming that the related $U(1)^\prime$ symmetry is broken spontaneously, the details of the scalar sector are not relevant for our purpose.} As can be seen from Fig.~\ref{fig:amuZp}, the contribution to $a_\mu$ decreases compared to the narrow-width limit, with the amount of the suppression depending on the mass of the decay products and the assumption for the high-energy behavior of the spectral function. 

\begin{figure}[t]
 \includegraphics[width=0.99\linewidth]{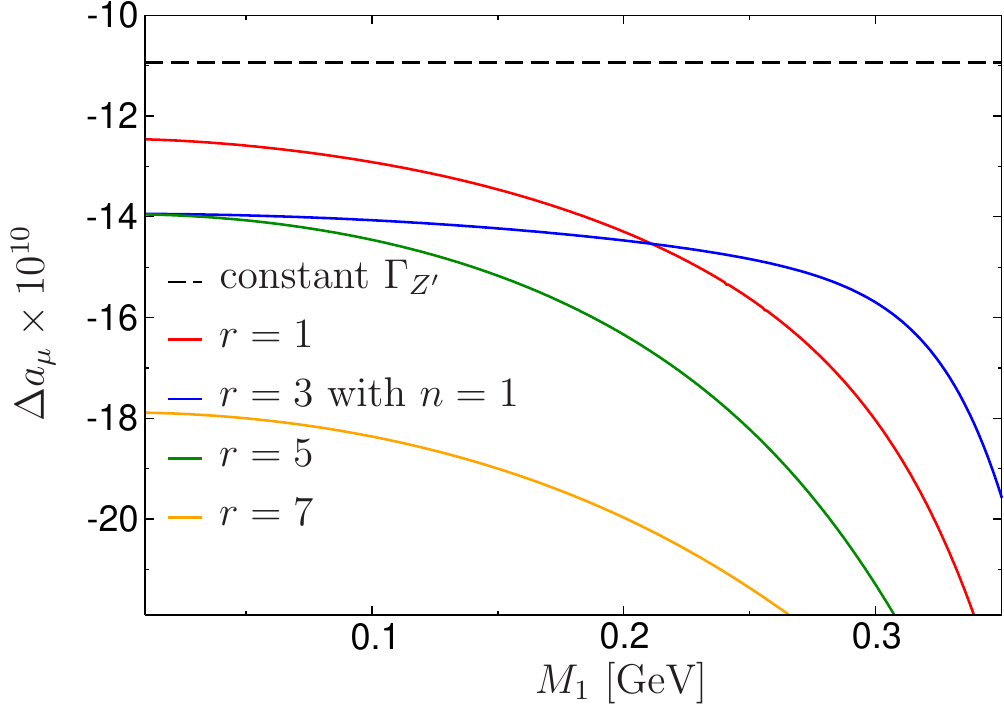}
 \caption{Shift $\Delta a_\mu$ due to a BSM effect in HVP for $M_{Z'}=0.8\GeV$, $\Gamma_{Z'}=0.2\GeV$, and $\eps_{Z'}=0.02$, as a function of $M_1$, for spectral functions with $\gamma_{Z'}(s)=(s-4M_1^2)^{r/2}/s$. 
 The case $r=3$, see Eq.~\eqref{gamma_spin_1}, requires $n=1$ 
  in Eq.~\eqref{xis}, while the other variants, illustrating the impact of modifying the threshold behavior, converge for $n=0$. The reference point for a constant width as in Eq.~\eqref{const_Gamma} (dashed line) should be treated with care due to the spurious imaginary parts below threshold.} 
 \label{fig:eepipi}
\end{figure}

Another possible application concerns the determination of the HVP contribution itself, given that the global $2.1\sigma$ tension between $e^+e^-$ data and the lattice-QCD calculation by BMWc~\cite{Borsanyi:2020mff} has now been confirmed at the level of about $4\sigma$ for the intermediate window in Euclidean time~\cite{RBC:2018dos} by several lattice collaborations~\cite{Colangelo:2022vok,Ce:2022kxy,ExtendedTwistedMass:2022jpw,FermilabLatticeHPQCD:2023jof,Blum:2023qou}. While independent lattice-QCD calculations of the entire HVP integral will require more time, some first conclusions can be drawn about the energy range in which changes to the $e^+e^-\to\text{hadrons}$ cross section would need to occur: first, the changes cannot occur too high in center-of-mass energy, otherwise serious tensions in the global electroweak fit would arise via the hadronic running of the fine-structure constant~\cite{Passera:2008jk,Crivellin:2020zul,Keshavarzi:2020bfy,Malaescu:2020zuc}, and this condition is confirmed by direct lattice-QCD calculations of the hadronic running itself~\cite{Borsanyi:2020mff,Ce:2022eix}. On the other hand, the deviation in the intermediate window shows that not all modifications can occur below $1\GeV$~\cite{Colangelo:2020lcg}, suggesting a more complicated pattern. In particular, already for the leading $2\pi$ channel large changes beyond the quoted experimental uncertainties would be required, and much larger relative changes for the subleading hadronic channels. 
In this situation, one could be inclined to entertain a BSM solution~\cite{DiLuzio:2021uty,Darme:2021huc}. However, arguably the most promising candidate, a broad $Z'$ interfering destructively with the SM signal, was shown to be excluded by other observables in Ref.~\cite{DiLuzio:2021uty}, assuming a modification of the $e^+e^-\to\pi^+\pi^-$ cross section according to
\beq
\label{const_Gamma}
\frac{\sigma_{\pi\pi}^\text{SM+NP}}{\sigma_{\pi\pi}^\text{SM}}=\bigg|1+\frac{\eps_{Z'} s}{s-M_{Z'}^2+i M_{Z'}\Gamma_{Z'}}\bigg|^2,
\eeq
where $\eps_{Z'}=g'^e_V(g'^u_V-g'^d_V)/e^2$ collects the $Z'$ couplings~\cite{DiLuzio:2021uty}. Since the interference before and after the resonance is opposite in sign (and cancels in the limit of a narrow resonance), the required net destructive interference mostly relies on the energy dependence of $\sigma_{\pi\pi}^\text{SM}$ and the kernel function in Eq.~\eqref{amu_HVP}. This effect can be enhanced by replacing the standard $Z'$ propagator via a KL representation. An example is shown in Fig.~\ref{fig:eepipi}, which illustrates how the asymmetry induced by increasing the threshold  or changing the functional form above can substantially increase
the change $\Delta a_\mu$ for the same set of couplings. 
We emphasize that the constraints derived in Ref.~\cite{DiLuzio:2021uty} remain severe (see App.~\ref{app:DeltaMpi2} for the pion mass difference and App.~\ref{app:LEP} for LEP bounds), so that one would likely have to push the $Z'$ mass beyond $1\GeV$ (including the effect on a variety of hadronic channels), introduce a large width, and tune the energy dependence of its spectral function to try and find a viable model.

\section{Conclusions}

In this work we presented a general framework how sizable width effects of new resonances can be consistently and efficiently incorporated into loop calculations. In particular, we showed how the underlying K\"all\'en--Lehmann representation can be matched to the Dyson series to constrain the properties of the spectral function from the perturbatively calculated one-loop self energy, capturing the leading two-loop effect in the case of a broad resonance. We calculated these spectral functions for several cases and showed how the calculation needs to be modified if new particles in one-loop diagrams acquire a large width, giving the general results for $a_\mu$ as an example. As a concrete application, we discussed how the effect of a broad $Z'$ could be included both directly in the calculation of $a_\mu$ and indirectly via its impact on the $e^+e^-\to\text{hadrons}$ cross section. Our results are applicable in quite general circumstances, reducing the calculation to the narrow-width limit for general masses together with a subsequent convolution with the spectral function, and can therefore be applied to a wide class of processes.

\begin{acknowledgments}
We thank Bastian Kubis, Luca Di Luzio, and Massimo Passera for helpful discussions. Financial support from  the SNSF (Project Nos.\ PP00P21\_76884 and  PCEFP2\_181117) is gratefully acknowledged. This work was performed in part at the Aspen Center for Physics, which is supported by National Science Foundation grant PHY-1607611.
\end{acknowledgments}

\appendix

\section{Dyson series and KL representation}
\label{app:KL}

Writing the one-particle-irreducible self energy of a scalar particle with bare mass $M_0$ as $\Sigma(p^2)$, the resummation of all self-energy insertions into a geometric series gives
\beq
i\Delta_D(p^2)=\frac{i}{p^2-M_0^2+\Sigma(p^2)}.
\eeq
To recast this resummation into a KL representation~\eqref{KL}, one can proceed as follows: 
absorbing the real part of the self energy into the mass renormalization (up to quadratic corrections) by means of a $Z$ factor defined by
\beq
p^2-M_0^2+\Re\Sigma(p^2)=Z^{-1}\big(p^2-M^2\big)+\Order\big[(p^2-M^2)^2\big],
\eeq
one finds the well-known expression
\beq
\label{resum}
i\Delta_D(p^2)=\frac{i Z}{p^2-M^2+i Z \,\Im\Sigma(p^2)}
\eeq
for the resummed propagator in terms of the on-shell renormalized mass $M$, which is valid up to higher orders in the expansion around the pole.\footnote{At higher orders in perturbation theory additional subtleties arise in the definition of the resonance parameters, see Refs.~\cite{Sirlin:1991fd,Passera:1996nk,Passera:1998uj} for the $W$ and $Z$ propagators in the SM.} Equation~\eqref{resum} reproduces a BW parameterization~\cite{Breit:1936zzb} once the energy-dependent width $\Gamma(s)$ is identified via $Z\,\Im\Sigma(s)=\sqrt{s}\Gamma(s)$. The key idea in deriving the spectral function from such BW parameterizations is that, while the analytic properties of the resummed result~\eqref{resum} are at odds with the KL representation~\eqref{KL}, the imaginary part does provide a useful approximation as long as the resonance does not become too broad.  Matching the imaginary parts then gives
\beq
\label{rho_scalar}
\rho_\phi(s)=\frac{Z}{\pi}\frac{\sqrt{s}\,\Gamma(s)}{\big(s-M^2\big)^2+s \big[\Gamma(s)\big]^2},
\eeq
where $Z$ can be determined from the normalization condition $\int ds\,\rho_\phi(s)=1$. In particular, $\rho_\phi(s)$ vanishes below the first threshold $s_\text{th}$ starting at which a non-vanishing imaginary part is generated in the self energy, as required by analyticity, and the correct threshold behavior is inherited from $\Im \Sigma(s)$ as well. In the limit of a narrow resonance, $\Gamma(s)=\Gamma\to 0$, Eq.~\eqref{rho_scalar} collapses to a $\delta$ function,
\beq
\rho_\phi(s)\to \delta\big(s-M^2\big),
\eeq
where we used that the normalization determines $Z=1$ in this case, and Eq.~\eqref{KL} reduces to the free propagator with mass $M$. 

In the spin-$1/2$ case, the KL representation generalizes to Eq.~\eqref{KL12}. To derive expressions for the spectral functions $\rho_{1/2}(s)$ from the imaginary part of the self energy we again start from the Dyson series
\beq
i\Delta_D(p)=\frac{i}{\slashed{p}-m_0+\slashed{p}\Sigma_1(p^2)+\Sigma_2(p^2)},
\eeq
where we separated the self energy, $\Sigma(p)=\slashed{p}\Sigma_1(p^2)+\Sigma_2(p^2)$ in analogy to the spectral function. Up to higher orders in the expansion around the renormalized mass $m$ this gives
\beq
i\Delta_D(p)=\frac{i Z}{\slashed{p}-m+iZ(\slashed{p}\,\Im\Sigma_1(p^2)+\Im \Sigma_2(p^2))},
\eeq
where the narrow-width limit 
suggests the identification of an energy-dependent width $\Gamma(p)=\slashed{p}\Gamma_1(p^2)+\Gamma_2(p^2)$ with $\Gamma_1(s)=2Z\,\Im\Sigma_1(s)$, $\Gamma_2(s)=2Z\,\Im \Sigma_2(s)$. Neglecting higher orders in the $\Gamma_i(s)$, this gives the matching relations 
\begin{align}
\label{spin12}
\rho_1(s)&=\frac{Z}{\pi}\frac{\frac{s+m^2}{2}\Gamma_1(s)+m\Gamma_2(s)}{\big(s-m^2\big)^2+s \big[\Gamma(s)\big]^2},\notag\\
\rho_2(s)&=\frac{Z}{\pi}\frac{m s\Gamma_1(s)+\frac{s+m^2}{2}\Gamma_2(s)}{\big(s-m^2\big)^2+s \big[\Gamma(s)\big]^2},\notag\\
s \big[\Gamma(s)\big]^2&=\frac{s+m^2}{2}\Big(s\big[\Gamma_1(s)\big]^2+\big[\Gamma_2(s)\big]^2\Big)\notag\\
&+2ms \Gamma_1(s) \Gamma_2(s).
\end{align}
In particular, in the limit $s=m^2$ one can read off $\Gamma = m \Gamma_1(m^2)+\Gamma_2(m^2)$ to make the identification with the constant width of the resonance. The positivity constraints translate to
\begin{align}
 \frac{s+m^2}{2}\Gamma_1(s)+m\Gamma_2(s)\geq 0,\notag\\
 \sqrt{s}\Gamma_1(s)-\Gamma_2(s)\geq 0,
\end{align}
and $Z$ can again be determined from the normalization condition for $\rho_1(s)$. For $\Gamma(s)=\Gamma\to 0$, Eq.~\eqref{spin12} reduces to
\beq
\rho_1(s)\to \delta\big(s-m^2\big),\qquad 
\rho_2(s)\to m\delta\big(s-m^2\big),
\eeq
where $Z=1$ follows from the normalization of $\rho_1(s)$. In the narrow-width limit one thus has 
$\sqrt{s}\rho_1(s)-\rho_2(s)= 0$ and Eq.~\eqref{KL12} reduces to the free propagator with mass $m$.

For the case of chiral fermions, Eq.~\eqref{KL12} generalizes to 
\beq
\label{KL12_chiral}
\Delta_F(p)=\int_0^\infty ds\, \frac{\slashed{p}\big[P_R\rho_1^R(s)+P_L\rho_1^L(s)\big]+\rho_2(s)}{p^2-s+i\eps},
\eeq
where the spectral functions follow from the matching relations
\begin{align}
\label{spin12_chiral}
\rho_1^{R/L}(s)&=\frac{Z}{\pi}\frac{\frac{s+m^2}{2}\bar\Gamma_1(s)\mp\frac{s-m^2}{2}\Delta\Gamma_1(s)+m\Gamma_2(s)}{\big(s-m^2\big)^2+s \big[\Gamma(s)\big]^2},\notag\\
\rho_2(s)&=\frac{Z}{\pi}\frac{m s\bar\Gamma_1(s)+\frac{s+m^2}{2}\Gamma_2(s)}{\big(s-m^2\big)^2+s \big[\Gamma(s)\big]^2},\notag\\
s \big[\Gamma(s)\big]^2&=\frac{s+m^2}{2}\Big(s\big[\bar\Gamma_1(s)\big]^2+\big[\Gamma_2(s)\big]^2\Big)\\
&+2ms \bar\Gamma_1(s) \Gamma_2(s)
+\frac{s(s-m^2)}{2}\big[\Delta\Gamma_1(s)\big]^2,\notag
\end{align}
with
\begin{align}
\bar\Gamma_1&=\frac{1}{2}\Big(\Gamma_1^R(s)+\Gamma_1^L(s)\Big),\notag\\ \Delta\Gamma_1&=\frac{1}{2}\Big(\Gamma_1^R(s)-\Gamma_1^L(s)\Big),
\end{align}
decomposing $\Sigma_1(s)$ and $\Gamma_1(s)$ into their left- and right-handed components accordingly. This results shows that, as long as $\Gamma_1$ is replaced by the mean of left- and right-handed contributions, the chirality difference enters always suppressed by a kinematical factor $s-m^2$, so that its effect vanishes on the resonance. For most situations with moderate widths in which a KL representation with spectral functions reconstructed from the self energy applies, it should therefore be sufficient to work with Eq.~\eqref{KL12}.

By the same argument, we can reexamine the relations~\eqref{spin12}. If corrections that scale with $(s-m^2)^2$ are ignored,
e.g., 
\begin{align}
 &\frac{s+m^2}{2}\Big(s\big[\Gamma_1(s)\big]^2+\big[\Gamma_2(s)\big]^2\Big)+2ms \Gamma_1(s) \Gamma_2(s)\notag\\
 &=m \sqrt{s}\big[\sqrt{s}\Gamma_1(s)+\Gamma_2(s)\big]^2\notag\\
 &+\frac{s\big[\Gamma_1(s)\big]^2+\big[\Gamma_2(s)\big]^2}{2(\sqrt{s}+m)^2}\big(s-m^2\big)^2,
\end{align}
the spectral functions simplify to 
\begin{align}
\label{spin12_simp}
\sqrt{s}\rho_1(s)&=
\rho_2(s)=\frac{Z}{\pi}\frac{m\sqrt{s}\Gamma(s)}{(s-m^2)^2+m \sqrt{s}\big[\Gamma(s)\big]^2},\notag\\
 \Gamma(s)&=\sqrt{s}\Gamma_1(s)+\Gamma_2(s),
\end{align}
where again the constant width is identified as $\Gamma=\Gamma(m^2)=m \Gamma_1(m^2)+\Gamma_2(m^2)$, and both positivity constraints are automatically fulfilled as long as $\Gamma(s)\geq 0$.    

\section{Self energies}
\label{app:self}

For the decay of a scalar particle $\phi$ according to
\beq
\Lagr_\phi=A_\phi \phi \phi_1\phi_2^\dagger + \bar F_1\big(C_S^\phi+ C_P^\phi\gamma_5) F_2 \phi+\text{h.c.},
\eeq
the imaginary part of the self energy is given by
\begin{align}
\label{Sigma_scalar}
\Im\Sigma^\phi&=\frac{q}{8\pi\sqrt{s}}\bigg(4\big(C_S^\phi\big)^2\big[s-(m_1+m_2)^2\big]\notag\\&+4\big(C_P^\phi\big)^2\big[s-(m_1-m_2)^2\big]+ 2A_\phi^2\bigg),
\end{align}
with center-of-mass momentum
\beq
q=\frac{\lambda^{1/2}\big(s,m_1^2,m_2^2\big)}{2\sqrt{s}}
\eeq
for the decay into particles with masses $m_1$, $m_2$, and step functions $\theta(s-s_\text{th})$, $s_\text{th}=(m_1+m_2)^2$ implied in Eq.~\eqref{Sigma_scalar} and the following. Since we are mainly interested in the energy dependence, as input for Eq.~\eqref{rho_scalar_final}, we only give a minimal variant of Eq.~\eqref{Sigma_scalar} for distinguishable complex scalars $\phi_1$, $\phi_2$ and Dirac fermions $F_1$, $F_2$; symmetry factors may need to be added for other cases.  

For the fermion Lagrangian
\begin{align}
\Lagr_F&=\bar F\big[C_V'\gamma^\mu+C_A'\gamma^\mu\gamma_5\big]F'X_\mu'\notag\\
&+\bar F\big[C_S'+C_P'\gamma_5\big]F'\phi'+\text{h.c.}, 
\end{align}
we get
\begin{align}
 \Im\bar\Sigma_1^F&=\frac{q}{8\pi\sqrt{s}}\bigg(C_{V}'^2\bigg[1+\frac{(\sqrt{s}-m_{F'})^2}{2M_{X'}^2}\bigg]\frac{s+m_{F'}^2-M_{X'}^2}{s}\notag\\
 &+C_S'^2\frac{s+m_{F'}^2-M_{\phi'}^2}{2s}\bigg)\notag\\
 &+\Big\{C_V'^2\to C_A'^2, C_S'^2\to C_P'^2, m_{F'}\to - m_{F'}\},\notag\\
  \Im\Sigma_2^F&=\frac{q}{8\pi\sqrt{s}}\bigg(C_V'^2\bigg[\frac{(\sqrt{s}-m_{F'})^2}{M_{X'}^2}-4\bigg]+C_{S}'^2\bigg)m_{F'}\notag\\
 &+\Big\{C_V'^2\to C_A'^2, C_S'^2\to C_P'^2, m_{F'}\to - m_{F'}\},
\end{align}
where the spin-$1$ contribution has been evaluated in Feynman gauge together with the appropriate Goldstone-boson contributions. We checked explicitly in general $R_\xi$ gauge that the linear combination $\sqrt{s}\Im \bar\Sigma_1^F(s)+\Im\Sigma_2^F(s)$, which enters in Eq.~\eqref{spin12_simp}, is indeed gauge invariant. 

Finally, for the decay of a spin-$1$ particle we use the Lagrangian 
\begin{align}
\Lagr_X&=iC_X \big(\phi_1\partial_\mu\phi_2^\dagger-\phi_2^\dagger\partial_\mu\phi_1\big)X^\mu\notag\\
&+\bar F_1\big(C_V^X\gamma^\mu+C_A^X\gamma^\mu\gamma_5\big)F_2X_\mu+\text{h.c.},
\end{align}
leading to 
\begin{align}
 \Im \Sigma^X&=\frac{q}{8\pi\sqrt{s}}\bigg(\frac{8}{3}q^2 C_X^2+
 8m_1m_2 \Big[\big(C_V^X\big)^2-\big(C_A^X\big)^2\Big]\notag\\
&+4\Big[\big(C_V^X\big)^2+\big(C_A^X\big)^2\Big]\Big(s-m_1^2-m_2^2-\frac{4}{3}q^2\Big)\bigg),
\end{align}
where $\Sigma^X$ is defined by the coefficient of the $g^{\mu\nu}$ term, as required for the calculation in Feynman gauge. The spectral function for the Goldstone-boson contribution can be derived as for the scalar case above.   

For completeness, we also provide the explicit perturbative expressions for the partial decay widths as they arise in the narrow-width approximation
\begin{widetext}
\begin{align}
\label{Gamma_pert}
 \Gamma_{\phi\to\phi_1\phi_2}&=\frac{\lambda^{1/2}\big(M^2_\phi,M_1^2,M_2^2\big)}{8\pi M^3_\phi}A_\phi^2,\notag\\
 \Gamma_{\phi\to F_1 F_2}&=\frac{\lambda^{1/2}\big(M^2_\phi,m_1^2,m_2^2\big)}{4\pi M^3_\phi}\bigg(\big(C_S^\phi\big)^2\big[M^2_\phi-(m_1+m_2)^2\big]+\big(C_P^\phi\big)^2\big[M^2_\phi-(m_1-m_2)^2\big]\bigg),\notag\\
 \Gamma_{F\to F'\phi'}&=\frac{\lambda^{1/2}\big(m^2_F,m_{F'}^2,M_{\phi'}^2\big)}{16\pi m^3_F}
 \bigg(C_S'^2\big[(m_F+m_{F'})^2-M_{\phi'}^2\big]+C_P'^2\big[(m_F-m_{F'})^2-M_{\phi'}^2\big]\bigg),\notag\\
  \Gamma_{F\to F'X'}&=\frac{\lambda^{1/2}\big(m^2_F,m_{F'}^2,M_{X'}^2\big)}{16\pi m^3_FM_{X'}^2}\bigg(C_V'^2\bigg[\lambda\big(m^2_F,m_{F'}^2,M_{X'}^2\big)+3M_{X'}^2\Big((m_F-m_{F'})^2-M_{\phi'}^2\Big)\bigg]\notag\\&+C_A'^2\bigg[\lambda\big(m^2_F,m_{F'}^2,M_{X'}^2\big)+3M_{X'}^2\Big((m_F+m_{F'})^2-M_{\phi'}^2\Big)\bigg]\bigg),\notag\\
  \Gamma_{X\to\phi_1\phi_2}&=\frac{\lambda^{3/2}\big(M^2_X,M_1^2,M_2^2\big)}{24\pi M_X^5}C_X^2,\notag\\
  \Gamma_{X\to F_1F_2}&=\frac{\lambda^{1/2}\big(M_X^2,m_1^2,m_2^2\big)}{12\pi M_X^5}\bigg(\big(C_V^X\big)^2\Big(2M_X^2+(m_1+m_2)^2\Big)\Big(M_X^2-(m_1-m_2)^2\Big)\notag\\
  &+\big(C_A^X\big)^2\Big(2M_X^2+(m_1-m_2)^2\Big)\Big(M_X^2-(m_1+m_2)^2\Big)\bigg),
\end{align}
\end{widetext}
where again symmetry factors may need to be applied for cases other than distinguishable complex scalars $\phi_i$ and Dirac fermions $F_i$.

\section{General expressions for $\boldsymbol{(g-2)_\mu}$}
\label{app:amu}

General expressions for $a_\mu$ have been derived in Ref.~\cite{Leveille:1977rc} for new interactions of the form
\begin{align}
\Lagr&=\bar \mu\big[C_V \gamma^\mu + C_A \gamma^\mu\gamma_5\big] F X_\mu\notag\\
&+\bar \mu\big[C_S+C_P \gamma_5\big] F \phi
+ \text{h.c.},
\end{align}
including new fermions $F$, (pseudo-) scalars $\phi$, and (axial-) vectors $X_\mu$, with masses $m_F$, $M_\phi$, $M_X$, respectively. For completeness, we reproduce the result
\begin{align}
\label{amu_narrow}
 a_\mu&=\frac{-Q_F m_\mu^2C_V^2}{4\pi^2}\int_0^1\frac{dx}{ \Delta^{(1)}(x,m_F^2,M_X^2)}\notag\\
 &\qquad\times\bigg[x(1-x)\bigg(x+\frac{2\Delta}{m_\mu}\bigg)+\frac{x^2\Delta^2}{2M_X^2}\bigg(1-x+\frac{m_F}{m_\mu}\bigg)\bigg]\notag\\
 &+\frac{Q_{X}m_\mu^2C_V^2}{4\pi^2}\int_0^1\frac{dx}{\Delta^{(2)}(x,m_F^2,M_X^2)}\bigg[x^2\bigg(1-x+\frac{2\Delta}{m_\mu}\bigg)\notag\\
 &\qquad+\frac{x(1-x)\Delta^2}{2M_X^2}\bigg(x+\frac{m_F}{m_\mu}\bigg)\bigg]\notag\\
 &-\frac{Q_F m_\mu^2C_S^2}{8\pi^2}\int_0^1\frac{dx}{\Delta^{(1)}(x,m_F^2,M_\phi^2)}x^2\bigg(1-x+\frac{m_F}{m_\mu}\bigg)\notag\\
 &+\frac{Q_\phi m_\mu^2C_S^2}{8\pi^2}\int_0^1\frac{dx}{\Delta^{(2)}(x,m_F^2,M_\phi^2)}x(1-x)\bigg(x+\frac{m_F}{m_\mu}\bigg)\notag\\
 &+\Big\{C_V^2\to C_A^2, C_S^2\to C_P^2, m_F\to - m_F\}, 
\end{align}
where 
\begin{align}
\Delta^{(1)}(x,m_F^2,M^2)&=m_\mu^2x^2+(m_F^2-m_\mu^2)x+M^2(1-x),\notag\\
\Delta^{(2)}(x,m_F^2,M^2)&=m_\mu^2x^2+(M^2-m_\mu^2)x+m_F^2(1-x),\notag\\
\Delta&=m_F-m_\mu,
\end{align}
and
the charges $Q_F$, $Q_{X}$, $Q_\phi$ are given in conventions in which $Q_\mu=Q_F+Q_{X}=Q_F+Q_\phi=-1$. In Ref.~\cite{Leveille:1977rc} these relations were derived in unitary gauge for the (axial-) vector contribution, but in the form given in Eq.~\eqref{amu_narrow} it is straightforward to read off how the full result emerges from summing the spin-$1$ part in Feynman gauge and the additional (pseudo-) scalar contribution proportional to $\Delta^2$ (for $Q_X>0$ there are also mixed diagrams with both spin-$1$ and Goldstone-boson propagators). The matching of the couplings, $C_S^2=C_V^2\Delta^2/M_X^2$, can be derived independently by demanding that the tree-level $\mu$--$F$ scattering amplitude be gauge independent. 

As a first generalization we consider the case in which the fermion width is neglected. This gives
\begin{widetext}
\begin{align}
\label{amu_rho}
 a_\mu&=\frac{-Q_F m_\mu^2C_V^2}{4\pi^2}\int ds\int_0^1\frac{dx}{ \Delta^{(1)}(x,m_F^2,s)}\bigg[x(1-x)\bigg(x+\frac{2\Delta}{m_\mu}\bigg)\rho_X(s)+\frac{x^2\Delta^2}{2s}\bigg(1-x+\frac{m_F}{m_\mu}\bigg)\rho_G(s)\bigg]\\
 &+\frac{Q_{X}m_\mu^2C_V^2}{4\pi^2}\int ds\,dt\int_0^1 dx\,L_{st}^{(2)}(x,m_F^2)\bigg[x\bigg(1-x+\frac{3\Delta}{2m_\mu}\bigg)\rho_X(s)\rho_X(t)+\frac{x\Delta}{2m_\mu}\rho_X(s)\rho_G(t)
 \notag\\
 &\qquad+\frac{(1-x)\Delta^2}{2\sqrt{st}}\bigg(x+\frac{m_F}{m_\mu}\bigg)\rho_G(s)\rho_G(t)\bigg]\notag\\
 &-\frac{Q_F m_\mu^2C_S^2}{8\pi^2}\int ds\,\rho_\phi(s)\int_0^1\frac{dx}{\Delta^{(1)}(x,m_F^2,s)}x^2\bigg(1-x+\frac{m_F}{m_\mu}\bigg)\notag\\
 &+\frac{Q_\phi m_\mu^2C_S^2}{8\pi^2}\int ds\, dt\,\rho_\phi(s)\rho_\phi(t)\int_0^1 dx\,(1-x)\bigg(x+\frac{m_F}{m_\mu}\bigg)L_{st}^{(2)}(x,m_F^2)
 +\Big\{C_V^2\to C_A^2, C_S^2\to C_P^2, m_F\to - m_F\},\notag
\end{align}
\end{widetext}
where
\beq
L^{(2)}_{st}(x,m_F^2)=\frac{\log\frac{\Delta^{(2)}(x,m_F^2,s)}{\Delta^{(2)}(x,m_F^2,t)}}{s-t},
\eeq
and $\rho_X$, $\rho_G$, $\rho_\phi$ are the respective spectral functions ($\rho_G$ denotes the Goldstone-boson part). As a final step, we provide the general result when also a finite width in the fermion propagators is admitted:
\begin{widetext}
\begin{align}
\label{amu_rho_F}
 a_\mu&=\frac{-Q_F m_\mu C_V^2}{4\pi^2}\int ds\,dt\,du\int_0^1dx\Bigg\{(1-x)L_{tu}^{(1)}(x,s)\bigg[m_\mu(x-2)\rho_1^F(t)\rho_1^F(u)+2\bar\rho_{tu}\bigg]\rho_X(s)\notag\\
 &\qquad+\frac{x\Delta_t\Delta_u}{2s}\bigg[m_\mu (1-x)\rho_1^F(t)\rho_1^F(u)+\bar\rho_{tu}\bigg]L_{tu}^{(1)}(x,s)\rho_G(s)+\frac{\Delta_t\Delta_u}{2s}\Delta\rho_{tu}\Big[\bar\Delta^{(1)}_{tu}(s)L_{tu}^{(1)}(x,s)-x \Big]\rho_G(s)\Bigg\}\notag\\
 &+\frac{Q_{X}m_\mu C_V^2}{4\pi^2}\int ds\,dt\,du\int_0^1 dx\,L_{st}^{(2)}(x,u)\Bigg\{x\bigg[\frac{3}{2}\rho_2^F(u)-m_\mu\Big(x+\frac{1}{2}\Big)\rho_1^F(u)\bigg]\rho_X(s)\rho_X(t)+\frac{x\Delta_u}{2}\rho_1^F(u)\rho_X(s)\rho_G(t)
 \notag\\
 &\qquad+\frac{(1-x)\Delta_u^2}{2\sqrt{st}}\Big(m_\mu x\rho_1^F(u)+\rho_2^F(u)\Big)\rho_G(s)\rho_G(t)\bigg]\notag\\
 &-\frac{Q_F m_\mu C_S^2}{8\pi^2}\int ds\,\rho_\phi(s)\int dt\,du\int_0^1 dx\Bigg\{x\bigg[m_\mu (1-x)\rho_1^F(t)\rho_1^F(u)+\bar\rho_{tu}\bigg]L_{tu}^{(1)}(x,s) +\Delta\rho_{tu}\Big[\bar\Delta^{(1)}_{tu}(s)L_{tu}^{(1)}(x,s)-x \Big]
 \Bigg\}
 \notag\\
 &+\frac{Q_\phi m_\mu C_S^2}{8\pi^2}\int ds\, dt\,\rho_\phi(s)\rho_\phi(t)\int du\int_0^1 dx\,(1-x)\Big(m_\mu x\rho_1^F(u)+\rho_2^F(u)\Big)L_{st}^{(2)}(x,u)\notag\\
 &+\Big\{C_V^2\to C_A^2, C_S^2\to C_P^2, \rho_2^F\to - \rho_2^F\},
\end{align}
\end{widetext}
with fermionic spectral functions $\rho_1^F$, $\rho_2^F$, $\Delta_s=\sqrt{s}-m_\mu$, as well as
\begin{align}
 L^{(1)}_{tu}(x,s)&=\frac{\log\frac{\Delta^{(1)}(x,t,s)}{\Delta^{(1)}(x,u,s)}}{t-u},\notag\\
 \bar\Delta_{tu}^{(1)}(s)&=\frac{1}{2}\Big(\Delta^{(1)}(x,t,s)+\Delta^{(1)}(x,u,s)\Big),\notag\\
 \Delta\rho_{tu}&=\frac{\rho_1^F(t)\rho_2^F(u)-\rho_1^F(u)\rho_2^F(t)}{t-u},\notag\\
 \bar\rho_{tu}&=\frac{1}{2}\Big(\rho_1^F(t)\rho_2^F(u)+\rho_1^F(u)\rho_2^F(t)\Big).
\end{align}

\section{$\boldsymbol{\pi^0\gamma}$ contribution to HVP}
\label{app:pi0gamma}

In a vector-meson-dominance (VMD) picture, the cross section for $e^+e^-\to \pi^0\gamma$ can be expressed as
\begin{align}
\label{pi0gamma_VMD}
 \sigma(e^+e^-\to\pi^0\gamma)&=\frac{2\pi^2\alpha^3}{3}\bigg(1-\frac{M_{\pi^0}^2}{s}\bigg)^3\Big|F_{\pi^0\gamma^*\gamma}(s,0)\Big|^2,\notag\\
 \frac{F_{\pi^0\gamma^*\gamma}(s,0)}{F_{\pi^0\gamma^*\gamma}(0,0)}&=1+\frac{1}{2}\sum_{V=\rho,\omega}\frac{s}{M_V^2-s-iM_V\Gamma_V},\notag\\
 F_{\pi^0\gamma^*\gamma}(0,0)&=F_{\pi\gamma\gamma}=\frac{1}{4\pi^2 F_\pi},
\end{align}
which upon insertion into Eq.~\eqref{amu_HVP} produces $a_\mu^{\pi^0\gamma}\simeq 4.23\times 10^{-10}$, close to the full result $a_\mu^{\pi^0\gamma}\simeq 4.38(6)\times 10^{-10}$~\cite{Hoid:2020xjs}. In contrast, the asymptotic expansion from Ref.~\cite{Blokland:2001pb}, also based on a VMD model for $F_{\pi^0\gamma^*\gamma}(s,0)$, gives $a_\mu^{\pi^0\gamma}\simeq 0.37\times 10^{-10}$, an order of magnitude smaller than the correct result. The reason for this mismatch can be traced back to the analytic structure: the VMD model employed in Ref.~\cite{Blokland:2001pb} does provide a realistic description in the space-like region, but it does not properly include the imaginary part of the HVP function. Setting $M_\rho=M_\omega=M_V$, the VMD model produces
\begin{align}
\label{PiVMD}
 \bar \Pi(k^2)&=-\alpha^2 F_{\pi\gamma\gamma}^2 M_V^4\bigg[\bigg(\frac{M_V^2}{M_V^2-k^2}\bigg)^2I(k^2)-I(0)\bigg],\notag\\
 I(k^2)&=\int_0^1 dx\int_0^{1-x}dy\frac{y}{x M_{\pi^0}^2+y M_V^2-k^2 x(1-x)}.
\end{align}
 Extracting the imaginary part from $I(k^2)$,  this diagrammatic calculation reproduces Eq.~\eqref{pi0gamma_VMD} by means of Eq.~\eqref{ImPi_R}, but, crucially, with widths $\Gamma_V=0$. The resulting function $\bar \Pi(k^2)$ as defined by the VMD model, Eq.~\eqref{PiVMD}, therefore cannot be continued into the time-like region. However, good analytic properties are important for the correct evaluation of the loop integral, and it is therefore no surprise that the phenomenological value is severely underestimated. Indeed, plugging in $\bar \Pi(k^2)$ from Eq.~\eqref{PiVMD} into the space-like master formula~\eqref{amu_space_like} gives a value of $0.41\times 10^{-10}$, very close to the result quoted in Ref.~\cite{Blokland:2001pb}. As the derivation of Eq.~\eqref{amu_space_like} assumed the validity of the dispersion relation~\eqref{barPi_disp}, this illustrates how the evaluation of loop integrals can fail if the correct analytic structure is not respected. In this case, it is ultimately the narrow width of the $\omega$ meson that leads to a sizable enhancement of the $\pi^0\gamma$ contribution.      

\section{Pion mass difference}
\label{app:DeltaMpi2}
 
A key constraint on the couplings of a $Z'$ to the light quarks in Ref.~\cite{DiLuzio:2021uty} derives from the pion mass difference. In the SM, most of the difference~\cite{ParticleDataGroup:2022pth,Crawford:1990jc,Daum:2019dav}
\beq
\Delta M_\pi^2 = \mpic^2-\mpin^2=1.26116(13)\times 10^{-3}\GeV^2,
\eeq
comes from electromagnetic interactions, with strong isospin breaking suppressed by $(m_u-m_d)^2$. This QCD effect can be estimated as~\cite{Gasser:1983yg}
\begin{align}
 \Delta M_\pi^2\big|_\text{QCD}&=\frac{2l_7 \mpi^4}{\Fpi^2}\bigg(\frac{m_u-m_d}{m_u+m_d}\bigg)^2\notag\\
 &=0.024(13)\times 10^{-3}\GeV^2,
\end{align}
where we used $l_7=2.5(1.4)\times 10^{-3}$~\cite{Frezzotti:2021ahg} (in line with Refs.~\cite{Gasser:1983yg,Boyle:2015exm}). With $\mpic-\mpin|_\text{QED}=\{4.622(95),4.534(60\}\MeV$ obtained in Refs.~\cite{Gagliardi:2021vpv,Feng:2021zek}, respectively, we conclude that 
\begin{align}
\label{DeltaMpilattice}
 \Delta M_\pi^2\big|_{Z'}&
=\Delta M_\pi^2- \Delta M_\pi^2\big|_\text{QCD}
- \Delta M_\pi^2\big|_\text{QED}\\
&=\{-0.032(29),-0.008(21)\}\times 10^{-3}\GeV^2\notag
\end{align}
could still originate from a $Z'$ contribution. The biggest effect for $M_{Z'}\lesssim 1\GeV$ comes from the elastic contribution, which can be evaluated within the Cottingham approach~\cite{Cottingham:1963zz,Ecker:1988te,Bardeen:1988zw,Donoghue:1993hj,Baur:1995ig,Donoghue:1996zn,Cirigliano:2020dmx,Cirigliano:2021qko,Stamen:2022uqh}. Following Ref.~\cite{Stamen:2022uqh}, one has
\begin{align}
M_\pi^2\big|_\text{QED}^\text{el}&=\frac{\alpha}{8\pi}\int_0^\infty ds \big[F_\pi^V(-s)\big]^2\Big(4W+\frac{s}{\mpi^2}(W-1)\Big)\notag\\
&=1.3(3)\times 10^{-3}\GeV^2,
\end{align}
where $W=\sqrt{1+4\mpi^2/s}$, i.e., the elastic part saturates the observed mass difference within uncertainties. The analogous formula for the $Z'$ contribution reads  
\begin{align}
M_\pi^2\big|_{Z'}^\text{el}&=\frac{(g'^u_V-g'^d_V)^2}{32\pi^2}\int_0^\infty ds\frac{s}{s+M_{Z'}^2} \big[F_\pi^V(-s)\big]^2\notag\\
&\times\Big(4W+\frac{s}{\mpi^2}(W-1)\Big),
\end{align}
notably of the opposite sign as Eq.~\eqref{DeltaMpilattice}. Allowing for a $Z'$ contamination at the level of the uncertainty in Eq.~\eqref{DeltaMpilattice} produces bounds 
\beq
\big|g'^u_V-g'^d_V\big|\lesssim 0.05\ldots 0.08,\qquad  M_{Z'}=(0\ldots 1)\GeV,
\eeq
corroborating the estimate from Ref.~\cite{DiLuzio:2021uty}. The limit becomes weaker with increasing $Z'$ mass according to the decoupling with $1/(s+M_{Z'}^2)$,  but in the mass range interesting for $e^+e^-\to\text{hadrons}$ data it remains very stringent. 
 
\section{LEP bounds for $\boldsymbol{Z'}$ couplings}
\label{app:LEP}

Including, compared to Ref.~\cite{DiLuzio:2021uty}, the contribution from the SM $Z$ boson,  
we find for the 
modification of the $e^+e^-\to q\bar q$ cross section due to a light $Z'$ at LEP~\cite{ALEPH:2013dgf}
\begin{widetext}
\begin{align}
 \frac{\sigma^\text{SM+NP}_{qq}}{\sigma^\text{SM}_{qq}}=1+\frac{2g'^e_Vg'^q_V}{e^2 Q_q}\frac{1-\frac{g_V^e g_V^q}{4Q_qc_W^2s_W^2}\frac{s}{s-M_Z^2}}{1-\frac{g_V^e g_V^q}{2Q_q c_W^2s_W^2}\frac{s}{s-M_Z^2}+\frac{[(g_V^e)^2+(g_A^e)^2][(g_V^q)^2+(g_A^q)^2]}{(4Q_qc_W^2s_W^2)^2}\big(\frac{s}{s-M_Z^2}\big)^2},
\end{align}
\end{widetext}
where $s_W=\sin\theta_W$, $c_W=\cos\theta_W$, and the $Z$-boson couplings $g_{V,A}^e$, $g_{V,A}^q$ are given in the conventions of Ref.~\cite{ParticleDataGroup:2022pth} (no summation over quark flavors is implied). Assuming again at most a $1\%$ change in the cross section, the $Z$-boson contribution tends to weaken the limits, with $|\eps_{Z'}|\leq 3.3 (1.7)\times 10^{-3}$ changing to $|\eps_{Z'}|\leq 5.1 (5.9)\times 10^{-3}$ for $q=u$ ($q=d$) at $\sqrt{s}\simeq 200\GeV$ (for smaller center-of-mass energies probed at LEP the $Z$-boson contribution becomes more important and thus the limit weaker). Moreover, the measured cross sections are the sum of $q=u,d,s,c,b$, and therefore the limit is further diluted if the $Z'$ does not couple in a flavor-universal way, by a factor of $5$ if only the coupling to a single flavor is assumed.   
 
\bibliography{amu}

\end{document}